\documentclass{iau-JDSS}
\usepackage{graphicx}

\title[Gamma-rays from binaries] 
{Gamma-ray emission from pulsar/massive-star binaries}

\author[G.E. Romero]   
{Gustavo E. Romero$^{1,\;2,}$%
  \thanks{Member of CONICET}}

\affiliation{$^1$Instituto Argentino de Radioastronom\'{\i}a (IAR, CCT La Plata, CONICET), C.C.5, 
(1894) Villa Elisa, Buenos Aires, Argentina  \break email: romero@iar-conicet.gov.ar\\[\affilskip]
$^2$Facultad de Ciencias Astron\'omicas y Geof\'{\i}sicas, 
Universidad Nacional de La Plata, Paseo del Bosque s/n, 1900 La Plata, Argentina  \break email: romero@fcaglp.unlp.edu.ar}

\pubyear{2009}
\volume{Volume xxx}  
\pagerange{119--126}
\date{?? and in revised form ??}
\jname{Highlights of Astronomy, Volume xxx}
\editors{T.M. Belloni, M. M\'endez, C. Zhang, eds.}
\begin{document}

\maketitle

\begin{abstract}
I present a review of the main phenomenological properties at high energies related to massive gamma-ray binaries and I discuss some aspects of pulsar models for these objects.

\keywords{Radiation mechanisms: non-thermal, stars: neutron, stars: early-type, stars: winds, outflows, gamma rays: observations,
gamma rays: theory.}
\end{abstract}

\firstsection 
\section{Introduction}

Many X-ray binaries are formed by a pulsar (rapidly rotating neutron star) and a massive, hot star. In most of these systems the soft X-ray emission is originated in an accretion disk, which is truncated at some distance of the neutron star by the magnetic pressure. The matter is then channeled onto the surface of the compact object through closed magnetic field lines and it impacts onto the magnetic poles of the star, producing hard X-rays by Bremsstrahlung. It has been speculated for a long time that such systems might also produce non-thermal radiation by various mechanisms. For instance, Maraschi \& Treves (1981) suggested that the interaction between the relativistic pulsar wind with the stellar wind of a Be star might result in efficient particle (electron) acceleration up to relativistic energies, with the consequent output of inverse Compton (IC) radiation in the gamma-ray domain. Such a model was later developed in more detail by Tavani et al. (1994) and Tavani \& Arons (1997) for the pulsar/Be system PSR B1259-63.  

A different type of model was proposed in the mid 1980s for Cygnus X-3, thought at that time to be a high-energy gamma-ray source. For instance, Berezinskii et al. (1986) considered the irradiation of the stellar atmosphere by relativistic protons produced by the pulsar. The result would be a gamma-ray and neutrino source.  

Finally, at the end of the same decade, Cheng \& Ruderman (1989) showed that in accreting systems, when the accretion disk rotates faster than the neutron star, a gap is open in the magnetosphere of the pulsar, and a strong potential drop is established in a charge-depleted region. Protons from the star might be accelerated in this gap along closed field lines in such a way that they would impact onto the accretion disk. This model was applied to the classic pulsar/Be binary A0535+262 by Romero et al. (2001), and further developed by Orellana et al. (2007).

In recent years several high-mass gamma-ray binaries have been detected by imaging atmospheric Cherenkov telescopes (IACTs), confirming some of these early predictions and posing new theoretical problems. In the next section I describe the main characteristics of the detected systems. 

\section{Known gamma-ray binaries}\label{gamma-b}

High-energy emission has been detected from four confirmed massive binaries by IACTs and there is a fifth source that is almost surely a binary, but additional information must yet be gathered, especially about the orbital parameters, before the full confirmation of this latter source.  With the exception of Cygnus X-1, which is a well-known black hole candidate, the spectral energy distributions (SEDs) of these systems are rather similar, with non-thermal radio emission, a power-law, featureless spectrum at X-rays, soft gamma-ray spectra, and strong variability, modulated by the orbital period (see Dubus 2006 and Skilton et al. 2009 for SEDs). 

\subsection{PSR B1259-63}\label{PSR}

PSR B1259-63 / SS 2883 is a binary system containing a B2Ve donor star and a 47.7 ms radio pulsar orbiting it every 3.4 years, in a very eccentric orbit with $e=0.87$. No radio pulses are observed when the neutron star (NS) is behind the circumstellar disk (an effect due to free-free absorption). Very high-energy (VHE) gamma-rays are detected when the NS is close to the periastron or crosses the stellar disk (Aharonian et al. 2005a).

The VHE spectrum can be fitted with a power-law of soft index ($\Gamma\sim -2.7\pm0.2$\footnote{$F_{\gamma}\propto E^{\Gamma}$.}) and explained by IC up-scattering of stellar photons. The light curve shows significant variability and a puzzling behavior, that can be reproduced using variable adiabatic and IC losses, along with the changing geometry (Khangulyan et al. 2007). The energetic electrons are produced at the termination shock of the pulsar wind. The cold relativistic wind should also cool by IC interactions with the stellar field. Khangulyan et al. (2007) presents detailed predictions for {\sl Fermi} satellite that can be used to determine the Lorentz factor of the cold wind. 

PSR B1259-63 has been detected as a variable radio source (e.g. Johnston et al. 2005). Pulsed radio emission is most of the time at a level of a few mJy. Transient unpulsed optically thin ($\alpha=-0.6$\footnote{$F_{\nu}\propto E^{\nu}$.}) radio emission up to 50 mJy is measured close to periastron passage. No VLBI images of the radio source, located in the southern hemisphere, have been published.

\subsection{LS I +61 303}\label{LSI} 

LS I +61 303 is a binary system containing a B0.5Ve donor star and a compact object of unknown mass (upper limit of $\sim 5$ $M_{\odot}$) orbiting it every 26.5 days, in an eccentric orbit with $e\sim 0.5-0.7$. No radio pulses are observed . VHE gamma-rays are detected after periastron (Albert et al. 2006, Acciari et al. 2009). The flux is variable. There are marginal detections at phases 0.2-0.4, close to periastron (located at phase 0.23). The maximum flux is detected at phases $0.5-0.7$, with values of $\sim$16\% of the Crab Nebula flux. The strong orbital modulation shows that the emission is produced, and affected, by the interplay between the two stars. The source has also been detected by {\sl Fermi} satellite, with a strong, periodic flux (Abdo et al. 2009). The maximum, contrary to what happens at higher energies, occurs immediately after the periastron passage. The spectrum shows a cutoff at $\sim 6$ GeV, indicating that the relation with the TeV emission is complex.  

At radio wavelengths, LSI +61 303 is highly variable, periodic, and has been resolved. Dhawan et al.  (2006) obtained a series of VLBA
images, spaced 3 days apart, and covering the entire orbit. The images show extended radio structures, with the phase near
periastron having indeed an apparent elongation away from the primary star. Since no macroscopic relativistic velocities are measured, 
the authors identify this elongation as ``a pulsar wind nebula shaped by the anisotropic environment, not a jet''. However, the situation is far from clear (see Romero et al. 2007).

\subsection{HESS J0632+057}\label{HESS}

HESS J0632+057 is a high-energy point-like source coincident with a B0pe star and a variable X-ray source (Hinton et al. 2009). The broadband spectrum looks like that of LS I +61 303, but no gamma-ray variability has been detected so far (see, nonetheless Acciari et al. 2009). The energetics, at the estimated distance of 1.5 kpc, requires a power of $dE/dt\sim 10^{34 - 35}$ erg/s, similar to other gamma-ray binaries. Recently, Skilton et al. (2009) have detected the source as a non-thermal radio emitter. {\sl Fermi} observations can be used to detect variability and possible periodic behavior, as in LS I +61 303.

\subsection{LS 5039}\label{LS}

LS 5039 is a binary system containing an O6.5V((f)) donor and a compact object of unknown mass (upper limit of $\sim 5$ $M_{\odot}$, Casares et al. 2005) orbiting it every 3.9 days, in a slightly eccentric orbit with $e\sim 0.35$. No radio pulses are observed. Persistent jet-like features have been reported several times with no puzzling behavior (same jet direction) up to now, suggesting a microquasar nature (Paredes et al. 2000). There are, however, no traces of accretion disk and no typical accretion variability. At X-rays the emission is a power-law, likely of synchrotron nature (e.g. Khangulyan et al. 2008).  The variable TeV emission is modulated with the orbital period of the binary system (Aharonian et al. 2005b). The flux maximum occurs at the inferior conjunction of the compact object (Aharonian et al. 2006). 

\subsection{Cygnus X-1}\label{Cyg}

Cygnus X-1 is a massive X-ray binary  containing an O9.7Iab donor star and an accreting black hole of at least 10 solar masses. The orbital period is 5.6 days, and the orbit, circular. Persistent radio emission, sometimes resolved in a jet-like feature, reveals its microquasar nature. It is a system that seems to be intrinsically different from the previous ones.

\section{Models}\label{models}

\subsection{Colliding wind models}

In colliding wind models, the compact object is a pulsar with a relativistic wind. This wind interacts with the stellar wind, producing a shock at the termination of the wind and particles can be accelerated there. Modulation naturally results from the changing conditions along the orbit. These type of models can reproduce the spectrum and light curve of PSR B1259-63 (Khangulyan et al. 2007), and they have been applied to LSI +61 303 and LS 5039 as well (Dubus 2006). However, is far from clear whether in the case of these binaries the observed radio morphology, the SEDs and the time variability can be accommodated within such a framework (see Romero et al 2007, Bosch-Ramon et al. 2008). The main test for colliding wind models would be the detection of the features produced by the interaction of the relativistic monoenergetic pulsar wind with the stellar field.  

\subsection{Models with accretion}

In models with accretion the non-thermal emission is produced in a collimated outflow. Such an outflow might be unstable due to the complex interaction with the ambient medium. Both leptonic and hadronic jet models have been proposed for LS 5039 and LS I +61 303 (e.g. Romero et al. 2005, Bosch-Ramon et al. 2006, Paredes et al. 2006, Dermer \& Boettcher 2006). These binaries present an X-ray behaviour different from what is observed in classic accreting black holes and NS. One possibility is that the accretion regime is dissipationless (Bogovalov \& Kelner 2005) and the angular momentum is removed by the outflow. It is usually thought that accreting models in high-mass binaries require black holes, but NS with low magnetic field can produce powerful jets (e.g. Sguera et al. 2009). The magnetic field can decay in a few million years because of the contamination of the NS's crust by impurities contained in the accreting inflow. If the magnetic field is low, the dissipationless accretion disk can approach to the star to form a magnetic tower that would eject the inflow. The detection of weak quasi-periodic oscillations at X-rays can indicate the presence of a radiatively inefficient disk in these objects. Jets interacting with clouds might also explain rapid flares in gamma-rays, something that is difficult to account for in an extended radiative region as that expected in a colliding wind scenario. 

\section{Conclusions}

Neutron stars in massive binary systems seem, under some specific conditions, to be able to produce non-thermal radiation up to very high energies. The relativistic particles can either be accelerated in the termination shock of a relativistic cold wind or in outflows where shocks might also mediate a process of diffusive acceleration. Opacity constraints and geometry make the non-thermal radiation of these systems periodic over a wide range of energies. The models briefly discussed in this review should be refined in the light of forthcoming multiwavelength observations in order to get a coherent picture for each source.   

%



\begin{acknowledgments}
This work has been supported by  the Argentine Agency ANPCyT (PICT-2007-00848, BID 1728/OC-AR), CONICET, and the Ministerio de Educaci\'on y Ciencia (Spain) under grant AYA 2007-68034-C03-01, FEDER funds.
\end{acknowledgments}

\end{document}